# STUDY ON THE INJECTION OPTIMIZATION AND TRANSVERSE COUPLING FOR CSNS/RCS*

M.Y. Huang[#], S.Y. Xu, J. Qiu, S. Wang,
Institute of High Energy Physics, Beijing, China

*Abstract*

The injection system of the China Spallation Neutron Source uses H$^-$ stripping and phase space painting method to fill large ring acceptance with the linac beam of small emittance. The emittance evolution, beam losses, and collimation efficiency during the injection procedures for different injection parameters, such as the injection emittances, starting injection time, twiss parameters and momentum spread, were studied, and then the optimized injection parameters was obtained. In addition, the phase space painting scheme which also affect the emittance evolution and beam losses were simulated and the optimization range of phase space painting were obtained. There will be wobble in the power supply of the injection bumps, and the wobble effects were presented. In order to study the transverse coupling, the injection procedures for different betatron tunes and momentum spreads were studied.

## INTRODUCTION

The accelerators of China Spallation Neutron Source (CSNS) consist of a 1.6GeV rapid cycling synchrotron (RCS) and a 80MeV proton linac which can be upgraded to 250MeV [1-2]. The RCS accumulates $1.56 \times 10^{13}$ protons in two intense bunches and operates at a 25Hz repetition rate with an initial design beam power of 100kW and is capable of being upgraded to 500kW. It has a four-fold lattice with four long straight sections for the injection, extraction, RF and collimations. During the physics design of the injection system [3], in order to control the influence of the strong space charge effects which are the main causes for beam losses in CSNS/RCS, the phase space painting method of injecting the beam of small emittance from the linac into the large ring acceptance was used. For the effective injection beam, there will be wobble effect in the power supply which can affect the emittance evolution, collimation efficiency and beam losses. In addition, the injection procedures for different momentum spreads and betatron tunes can be simulated while studying the transverse coupling. Using ORBIT which is a particle tracking code for rings [4], one can simulate the injection procedure and study the emittance evolution, collimation efficiency and beam losses.

*Work supported by National Natural Science Foundation of China (No. Y211A3105C)
[#]huangmy@ihep.ac.cn

## OPTIMIZATION OF INJECTION PARAMETERS

The CSNS/RCS injection system is to inject the H$^-$ beam into the RCS with high precision and high transport efficiency. During the injection procedure, there are some factors, such as the betatron tunes, injection emittances, starting injection time, twiss parameters, momentum spread, chopping rate, space charge effect, phase space painting scheme, which can affect the emittance evolution, collimation efficiency and beam losses. Some works had been done for the injection optimization [5-6]. In this section, we discuss the injection parameters optimization due to the injection emittances, starting injection time, twiss parameters, momentum spread, and the optimization of phase space painting scheme will be discussed in the next section.

With the code ORBIT, the injection procedure of milti-turn particle tracking were simulated while the injection rms emittances between (0.1, 0.1) and (10.0, 10.0), starting injection time between -0.3ms and -0.09ms, momentum spread between 0.01% and 0.5%, alphas between (0.001, 0.001) and (1.0, 1.0). Table 1 shows the simulation results for different injection parameters. It can be found that while the injection rms emittances smaller than (1.0, 1.0), momentum spread smaller than 0.1%, alphas smaller than (1.0, 1.0), the collimation efficiency are larger than 89%, the beam losses are constrained smaller than 1%, and the rms emittances are constrained in a reasonable ranges. Therefore, the optimization ranges for the injection parameters can be obtained.

Table 1: Simulation Results for Different Injection Parameters

| Parameters Values | | Beam Losses | Collimation Efficiency | $(\varepsilon_x, \varepsilon_y)_{rms}$ (mm mrad) |
|---|---|---|---|---|
| $\Delta p/p$ | 0.01% | 0.81% | 93.5% | (33.4, 33.4) |
|  | 0.10% | 0.62% | 89.1% | (31.3, 31.9) |
| $t_{inj}$ (ms) | -0.3 | 0.75% | 91.6% | (34.1, 32.1) |
|  | -0.14 | 0.59% | 89.7% | (31.3, 31.9) |
| $(\alpha_x, \alpha_y)$ | (0.01, 0.01) | 0.67% | 90.3% | (31.9, 32.6) |
|  | (1.0, 1.0) | 0.71% | 90.2% | (32.6, 34.0) |
| $(\varepsilon_x, \varepsilon_y)$ | (0.1, 0.1) | 0.46% | 90.4% | (31.3, 30.5) |
|  | (1.0, 1.0) | 0.61% | 88.9% | (31.2, 32.9) |

## STUDY ON PHASE SPACE PAINTING

In the above section, the injection optimization for different parameters had been discussed. During the injection procedure, the phase space painting scheme can also affect the emittance evolution, collimation efficiency and beam losses. There were some works which had been done for the phase space painting optimization [7]. In this section, the injection simulation for different phase space painting ranges will be discussed.

Using the code ORBIT, the injection procedures for different phase space painting ranges were simulated and the emittance evolutions and beam losses were obtained. It can be found from Table 2 that the beam losses, collimation efficiency and the rms emittances all increase with the painting range of y axe. In addition, the transverse coupling between the x and y rms emittances was found from Figure 1. Therefore, the optimization range of phase space painting can be obtained.

Table 2: Simulation Results for Different Phase Space Painting Ranges

| Painting range of y axe (mm) | Beam Losses | Collimation Efficiency | $(\varepsilon_x, \varepsilon_y)_{rms}$ (mm mrad) |
|---|---|---|---|
| 28→0 | 0.31% | 85.4% | (30.9, 28.1) |
| 32→0 | 0.61% | 91.1% | (31.0, 33.1) |
| 36→0 | 1.90% | 92.8% | (33.4, 38.8) |
| 40→0 | 5.85% | 94.8% | (36.6, 44.6) |

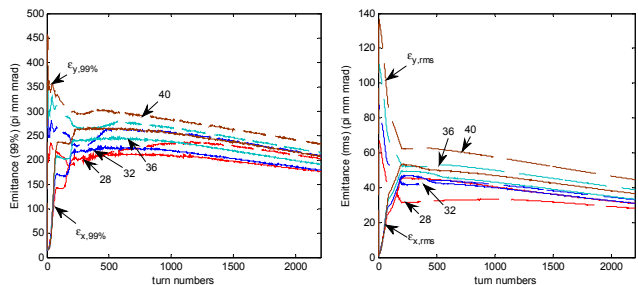

Figure 1: The emittance evolution for different phase space painting ranges.

## WOBBLE EFFECTS IN THE POWER SUPPLY

The phase space painting curve for CSNS/RCS was idealization which didn't considered the wobble of the power supply for the kickers in the above discussion [8]. However, the wobble effects in the power supply can affect the emittance evolution and beam losses during the injection procedure and need to be considered. In this section, the injection procedure with wobble in the power supply will be discussed in detail.

With the code ORBIT, the injection procedures for different wobble time of the power supply were simulated and the emittance evolutions and beam losses were obtained. It can be found from Table 3 that the beam losses and collimation efficiency both increase with the wobble time. The transverse coupling between x and y rms emittance evolutions can be found in Figure 2 and it disappear if the wobble time large enough. Therefore, the wobble in the power supply needs to be considered during the physics design of injection system for CSNS/RCS.

Table 3: Simulation Results for the Wobble Effects in the Power Supply

| Wobble Time (μs) | Beam Losses | Collimation Efficiency | $(\varepsilon_x, \varepsilon_y)_{rms}$ (mm mrad) |
|---|---|---|---|
| 0 | 0.90% | 90.3% | (30.8, 32.5) |
| 8 | 1.66% | 92.7% | (32.5, 31.5) |
| 14 | 3.12% | 93.4% | (32.6, 31.0) |
| 17 | 8.06% | 94.4% | (34.6, 29.6) |

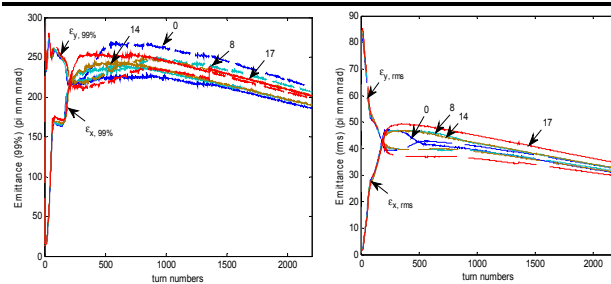

Figure 2: The emittance evolution for different wobble time of the power supply.

## TRANSVERSE COUPLING

In order to study the transverse coupling between x and y rms emittance evolution in more detail, the injection procedures for different momentum spreads and betatron tunes need to be discussed.

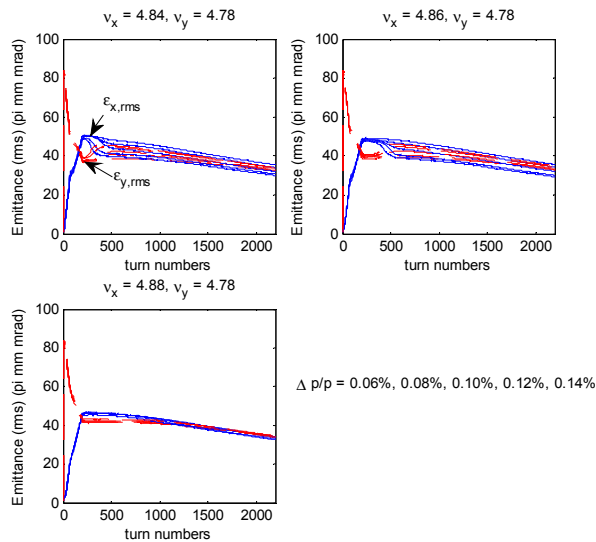

Figure 3: The emittance evolution for different momentum spreads and betatron tunes.

With the code ORBIT, the injection procedures for different momentum spreads and betatron tunes were simulated. Figure 3 and 4 show the x and y rms emittance evolution for different momentum spreads and betatron tunes. It can be found that the transverse coupling increases with the momentum spread and also depends on the betatron tunes. Therefore, the transverse coupling needs to be considered during the physics design of CSNS/RCS.

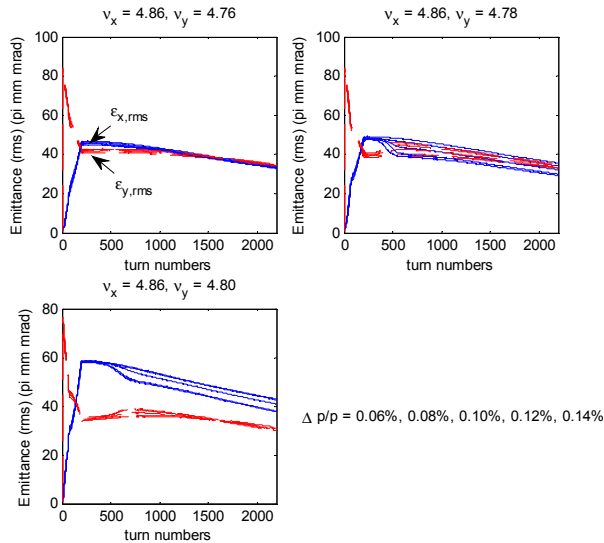

Figure 4: Similar to Figure 2 but for different betatron tunes.

## CONCLUSIONS

The injection procedures for different injection emittances, starting injection time, twiss parameters, momentum spread, and phase space painting ranges have been studied and the optimization ranges for the injection parameters were obtained. For the effective injection beam, the wobble effects in the power supply of the injection bumps have been discussed and it can be found that the beam losses and collimation efficiency both increase with the wobble time. In addition, the transverse coupling was found and it disappears if the wobble time large enough. Finally, the injection procedures for different momentum spreads and betatron tunes have been studied and it can be found that the transverse coupling increases with the momentum spread and also depends on the betatron tunes.

## ACKNOWLENDGMENTS

The authors want to thank CSNS colleagues for the discussion and consultations.